\documentclass[twocolumn]{aastex63}

\usepackage{float}
\usepackage{graphicx}

\hypersetup{linkcolor=red,citecolor=green,filecolor=cyan,urlcolor=magenta}

\newcommand{\radm}{\,rad m$^{-2}$} 

\received{XXX}
\revised{XXX}
\accepted{XXX}
\submitjournal{ApJ}

\shorttitle{The mode switching in pulsar J1326$-$6700}
\shortauthors{Z. G. Wen et al.}

\begin{document}

\title{The mode switching in pulsar J1326$-$6700}

\correspondingauthor{Z. G. Wen}
\email{wenzhigang@xao.ac.cn}

\author{Z. G. Wen}
\affiliation{Xinjiang Astronomical Observatory, Chinese Academy of Sciences, \\
150, Science-1 Street, Urumqi, Xinjiang, 830011, China}
\affiliation{Guizhou Provincial Key Laboratory of Radio Astronomy and Data Processing, \\
Guiyang, Guizhou, 550001, China}
\affiliation{Key laboratory of Radio Astronomy, Chinese Academy of Sciences, \\
Nanjing, 210008, China}

\author{W. M. Yan}
\affiliation{Xinjiang Astronomical Observatory, Chinese Academy of Sciences, \\
150, Science-1 Street, Urumqi, Xinjiang, 830011, China}
\affiliation{Key laboratory of Radio Astronomy, Chinese Academy of Sciences, \\
Nanjing, 210008, China}

\author{J. P. Yuan}
\affiliation{Xinjiang Astronomical Observatory, Chinese Academy of Sciences, \\
150, Science-1 Street, Urumqi, Xinjiang, 830011, China}
\affiliation{Key laboratory of Radio Astronomy, Chinese Academy of Sciences, \\
Nanjing, 210008, China}

\author{H. G. Wang}
\affiliation{School of Physics and Electronic Engineering, \\
Guangzhou University, 510006, Guangzhou, PR China }
\affiliation{Xinjiang Astronomical Observatory, Chinese Academy of Sciences, \\
150, Science-1 Street, Urumqi, Xinjiang, 830011, China}

\author{J. L. Chen}
\affiliation{Department of Physics and Electronic Engineering, \\
Yuncheng University, Yuncheng, Shanxi, 044000, China}

\author{M. Mijit}
\affiliation{School of Physical Science and Technology, \\
Xinjiang University, urumqi, Xinjiang, 830046, China}

\author{R.Yuen}
\affiliation{Xinjiang Astronomical Observatory, Chinese Academy of Sciences, \\
150, Science-1 Street, Urumqi, Xinjiang, 830011, China}

\author{N. Wang}
\affiliation{Xinjiang Astronomical Observatory, Chinese Academy of Sciences, \\
150, Science-1 Street, Urumqi, Xinjiang, 830011, China}
\affiliation{Key laboratory of Radio Astronomy, Chinese Academy of Sciences, \\
Nanjing, 210008, China}

\author{Z. Y. Tu}
\affiliation{Xinjiang Astronomical Observatory, Chinese Academy of Sciences, \\
150, Science-1 Street, Urumqi, Xinjiang, 830011, China}

\author{S. J. Dang}
\affiliation{Xinjiang Astronomical Observatory, Chinese Academy of Sciences, \\
150, Science-1 Street, Urumqi, Xinjiang, 830011, China}
\affiliation{School of Physics and Electronic Science, Guizhou Normal University, \\
No. 116, Baoshan Road, Yunyan Distric, Guiyang, 550001, China}

\begin{abstract}
We report on a detailed study on the mode switching in pulsar J1326$-$6700 
by analysis the data acquired from the Parkes 64-m radio telescope at 1369 MHz.
During the abnormal mode, the emission at the central and trailing components
becomes extremely weak. 
Meanwhile, the leading emission shifts towards earlier by almost 2 degrees, and
remains in this position for typically less than a minute.
The mean flux density of the normal mode is almost five times that of the
abnormal mode.
Our data show that 85\% of the time for PSR J1326$-$6700 was in the normal mode
and 15\% was in the abnormal mode.
The intrinsic distributions of mode timescales can be well described by Weibull
distributions, which presents a certain amount of memory in mode switching.
Furthermore, a quasi-periodicity has been identified in the mode switching in
pulsar J1326$-$6700.
The estimated delay emission heights based on the kinematical effects indicate
that the abnormal mode may be originated from higher altitude than the normal mode.
\end{abstract}

\keywords{stars: neutron -- pulsars: general -- pulsars: individual: PSR
J1326$-$6700}

\section{Introduction}
\label{sec:intro}
Pulsars are rapidly rotating, highly magnetized neutron stars which emit
radio electromagnetic radiation along their magnetic axes.
The discrete and regular pulses are received while our line of sight sweeps
across the radio beam as the star rotates.
Two primary single pulse modulations intrinsic to pulsar radio emission are the
phenomena of nulling and mode changing (or mode switching).
Nulling is the abrupt cessation of pulsed emission for several periods.
Mode changing is the sudden switches between two or more distinct emission patterns.
Both of these effects have been investigated extensively and are suggested to be
originated from large scale and persistent changes in the magnetospheric current
distribution \citep{Wang+etal+2007}.

Radio strong pulsar J1326$-$6700 is characterized by the intriguing combination of
three canonical pulse modulation phenomena: nulling, mode changing and
occasional subpulse drifting \citep{Wang+etal+2007}.
PSR J1326$-$6700 first attracted our attention because the average pulse profiles
for two modes present extremely distinct difference.
During the abnormal mode episode, sporadic emission appears at the leading edge
of the profile lasting for a dozen or more pulses.
Meanwhile, the emission from the usual window ceases.
The effect bears some resemblance to the anomalous emission events (recently
referred as `swooshes') presented in pulsars B1859+07 and B0919+06 
\citep{Rankin+etal+2006}, whose emission shifts to an earlier longitude gradually.
The possibility of conventional mode changing was ruled out because of the gradual
onsets and relaxations of the events.
Among several proposed explanations, the changes in emission altitude appears 
possible to explain the gradual nature of the event onsets and returns.
However, the displacement of the emitting region requires too large vertical
height in the magnetosphere which greatly exceed the height above the neutron
star's surface.

In this paper, by using the archived data observed with the Parkes 64-m radio
telescope at 1369 MHz, we focus on the specific characteristics of normal and
abnormal emission modes.
Furthermore, a perplexing periodicity is shown in the mode switching, which may
in turn open up a new potential avenue of interpretation.
Details of the observations are described in Section~\ref{sec:obs}.
Results are presented in Section~\ref{sec:results}.
The implication of our results to the emission geometry and possible mechanisms
for the mode switching are discussed in Section~\ref{sec:diss}.
Finally, Section~\ref{sec:con} summarizes the results and discussion.

\section{Observations}
\label{sec:obs}
The analyses in this paper are based on five available observations from the
Parkes pulsar data archive \citep{Hobbs+etal+2011}, all of which were 
carried out using the Parkes 64 meter radio telescope and the multi-beam receiver.
The data were recorded with one of the Parkes digital filterbank systems
(PDFB3/4) in incoherent mode.
Full-Stokes spectra were acquired using 512 channels across a 256 MHz passband
centered at 1369 MHz radio frequency.
The data were integrated in time for 256 $\rm \mu$s per spectrum before
recording to disk with 8-bit quantization.
In order to calibrate flux and polarization precisely, pulsed noise signals were
linearly injected into the feed horn prior to the first-stage low-noise
amplifier.
The system equivalent flux density (SEFD) on cold sky was determined via paired
observations of the unpolarized extragalaxy 3C218 (Hydra A) with an assumed flux
at 1400 MHz of 43.1 Jy.
The detailed information of the observations are listed in Table~\ref{tab:obs}.

\begin{table*}[h]
    \centering
    \caption{Summary of radio observations of PSR J1326$-$6700.}
    \label{tab:obs}
	\begin{tabular}{ccccccc}
        \hline 
        \hline
		Date & Start time & Backend & SEFD & Duration & Flux density & RM \\
		(yyyy$-$mm$-$dd) & (UTC) & & (Jy) & (s) & (mJy) & (\radm) \\
        \hline
		2012$-$01$-$15 & 20:14:24 & PDFB4 & 37.1 & 1505 & $11.26\pm0.02$ & $-46.6\pm0.9$ \\
		2014$-$05$-$25 & 03:36:17 & PDFB4 & 35.7 & 7204 & $9.25\pm0.01$  & $-56.3\pm0.7$ \\
		2014$-$05$-$30 & 06:04:42 & PDFB4 & 35.6 & 6384 & $10.92\pm0.02$ & $-52.3\pm0.8$ \\
		2014$-$10$-$15 & 22:07:09 & PDFB3 & 35.3 & 7205 & $22.99\pm0.03$ & $-25\pm1$ \\
		2014$-$12$-$02 & 20:38:09 & PDFB4 & ...  & 1466 & ...  & ... \\
        \hline
    \end{tabular}
	\begin{flushleft}
		\textbf{Notes.} No paired calibrator was observed for the 2014$-$12$-$02
		observation.
		The paired calibrator for the 2014$-$10$-$15 observation was recorded
		with the PDFB4, whereas the target source was recorded with the PDFB3.
		Therefore, the great deviations of flux density and RM are the result of
		equipment difference.
	\end{flushleft}
\end{table*}

The PSRCHIVE\footnote{https://psrchive.sourceforge.net/} \citep{Hotan+etal+2004} and 
DSPSR\footnote{http://dspsr.sourceforge.net/} \citep{Straten+Bailes+2011} pulsar 
analysis software packages were used in off-line data reduction.
Initially, the single pulse integrations were obtained by folding the data into
1024 phase bins per pulse period.
The pulsar's rotational ephemerides was taken from the ATNF pulsar catalogue
V1.61\footnote{http://www.atnf.csiro.au/research/pulsar/psrcat/}
\citep{Manchester+etal+2005}.
The radio frequency interference (RFI) was then mitigated automatically in
the frequency domain using median filtering technique and 5 per cent of each
band edge were zero weighted.
Subsequently, the data were calibrated to compensate for instrumental gain and
phase variations across the band, converted to Stokes parameters, and placed on
a flux density scale.
From the resulting multi-frequency polarization profiles, the rotation measure
(RM) was obtained by brute-force search for peak linear polarization and then
iterative refinement of differential position angle following the method
described by \citet{Han+Manchester+2006}.
The cause of significant differences in the RM value of PSR J1326$-$6700 from the
different observations is not clear.
Finally, the full-Stokes individual pulses were obtained by removing the
dispersive smearing between sub-channels, which provide the basis for the
analyses described in the following sections.



\section{Results}
\label{sec:results}

\subsection{Pulse sequence dynamics}
\begin{figure}[h]
	\centering
	\includegraphics[width=8.0cm,height=16.0cm,angle=0]{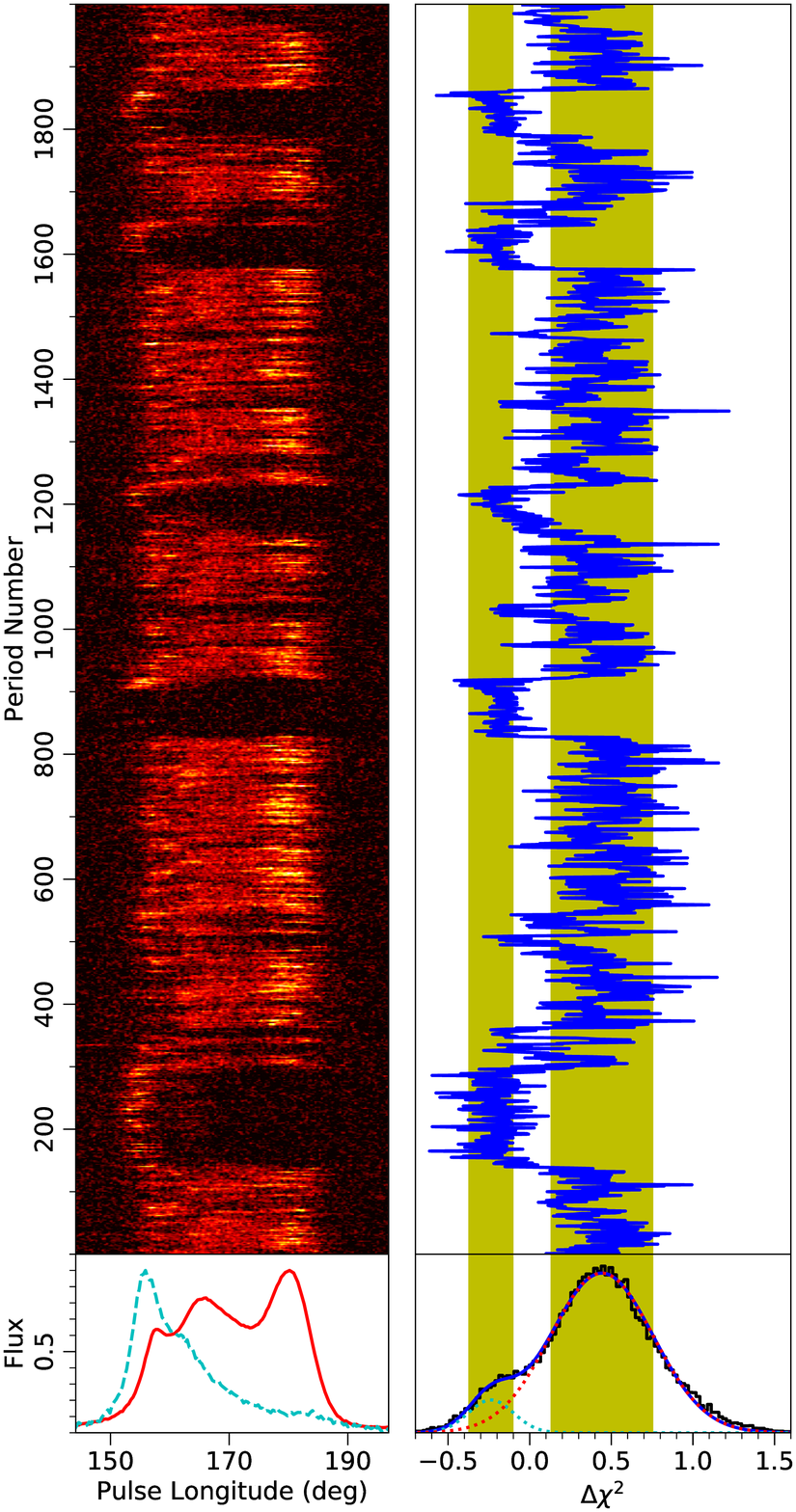}
	\caption{Top-left panel: longitude-time diagram showing intensity variations for
	2000 individual pulses from PSR J1326$-$6700.
	Time extends from left to right through pulses and successive pulses are
	plotted from bottom to top.
	Bottom-left panel: the comparisons of the mean pulse profile for the normal
	emission mode (solid red) and abnormal emission mode (dashed cyan).
	Top-right panel: the $\Delta \chi^2$ for the corresponding profiles on the
	left.
	Bottom-right panel: histogram of the $\Delta \chi^2$ for the entire data
	set is indicated with the black solid line.
	The blue solid line represents the fitting based on the combination of two
	Gaussian components (dashed cyan and dashed red lines).
	The yellow bars indicate the $\pm1\sigma$ range around the fitted two peaks.}
	\label{pic:sgl_demo}
\end{figure}

With the benefit of Parkes high sensitivity, PSR J1326$-$6700 is bright enough
to study its single pulse sequences, and a color-coded pulse sequence displays 
several of the pulsar’s emission behaviors in the top-left panel of 
Figure~\ref{pic:sgl_demo}.
An intriguing combination of nulling and mode changing is clearly shown.
The pulses with no detectable emission from central and trailing components are 
clearly visible.
During the partial nulls, an obvious abnormal mode emission appears to flicker
on at the leading edge the profile for typically less than a minute.
The sporadic emission in the abnormal mode is frequently separated by short nulls.
On several occasions, the two modes are separated by a complete null interval where 
the coherent radio emission in the whole pulse period ceases.

\subsection{Pulse energy distribution}
\begin{figure}[h]
	\centering
	\includegraphics[width=8.0cm,height=6.0cm,angle=0]{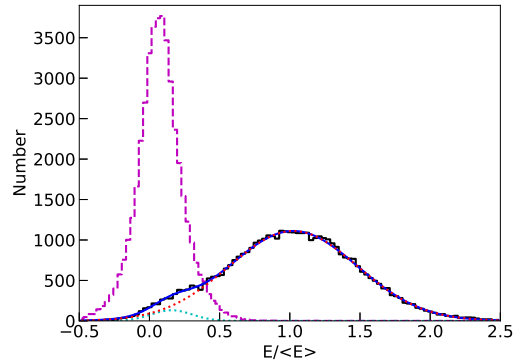}
	\caption{Pulse energy distributions for the on-pulse (black solid histogram)
	and off-pulse (magenta dashed histogram) regions are calculated from all
	observations.
	The energies are normalized by their respective mean on-pulse energies.
	The blue solid line represents the fitting for the on-pulse energy
	distribution based on the combination of two Gaussian components (cyan and
	red dotted lines) with the R-square of 0.997.
	}
	\label{pic:energy}
\end{figure}

In order to characterize the pulse nulling properties statistically, the pulse
energy distributions for the on-pulse region and off-pulse region with similar 
lengths, after a normalisation by the mean pulse energy, are presented in
Figure~\ref{pic:energy}.
The off-pulse energy histogram, centering around zeros, represents a Gaussian 
random noise contributed by the telescope noise.
While the on-pulse energy distribution shows the presence of two distinct
regions, which corresponds to bistable emission modes.
The two peaks as determined from a fit with the sum of two normal distributions
are obviously larger than zero, which implies that the apparent nulls are in an
abnormal emission mode rather than real null pulses.

\subsection{Identification of mode switching}
The single pulse energy distribution indicates the existence of mode switching
in PSR J1326$-$6700, with the pulse profile switching on timescales of seconds
between two quasi-stable modes, which differ in intensity.
In order to determine mode switching properties such as the timescale and
polarization, a quantitative metric is defined to distinguish whether an
individual pulse belongs to the normal or the abnormal emission mode, which is
given by \citep{Mahajan+etal+2018}
\begin{equation}
	\centering
	\rm
	\Delta \chi_i^2 =
	\sum_{\phi}\frac{(p_i(\phi)-p_a(\phi))^2-(p_i(\phi)-p_n(\phi))^2}{\sigma(\phi)^2},
\end{equation}
where $p_i(\phi)$ is an individual pulse profile, $p_n(\phi)$ and $p_a(\phi)$
are the average pulse profiles for normal and abnormal modes respectively, and
$\sigma(\phi)$ is the standard deviation from the mean of the entire data set.
The mode metric $\Delta \chi^2$ corresponding to individual pulses are indicated
in the top-right panel of Figure~\ref{pic:sgl_demo}.

A histogram of mode metric for our whole data set of $\sim$ 44 thousand pulses is
shown in the bottom-right panel of Figure~\ref{pic:sgl_demo}, which clearly
shows a bi-modal distribution.
We fit the histogram with a sum of two normal distributions, finding centers for
the normal and abnormal mode of $\Delta \chi^2 = 0.442\pm0.003$ and
$-0.238\pm0.009$, respectively, and associated widths of $\sigma_{\Delta \chi^2}
= 0.315\pm0.004$ and $0.138\pm0.009$.
The $\Delta \chi^2$ distribution of the normal mode is considerably wider than
that of the abnormal mode, which may indicate that the abnormal mode is more
stable than the normal mode, and the timescale for the normal mode to get stable
is longer than that for the abnormal mode.

It is noted that some contamination by mode transitions is clearly presented, and 
it is not always obvious to which mode an individual pulse belongs due to low
signal-to-noise ratio (S/N).
As suggested that the durations of transitions do not necessarily represent the
true time spans for which mode changes occur, a $\Delta \chi^2$ threshold is
determined as the center of the inner 1$\sigma$ boundaries of normal and
abnormal modes.
Thus, all of the individual pulses are associated with either the normal or
the abnormal mode.

\begin{figure}[h]
	\centering
	\includegraphics[width=8.0cm,height=8.0cm,angle=0]{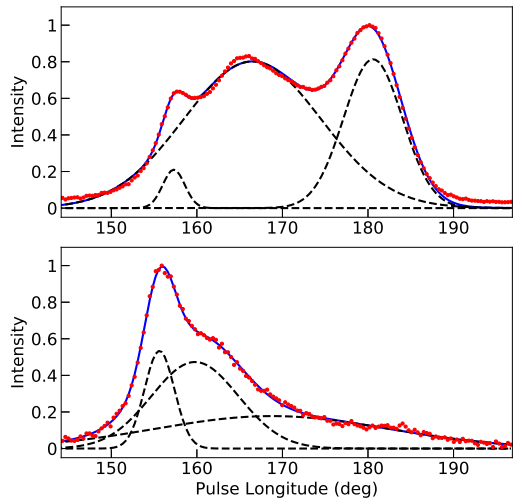}
	\caption{Observed pulse profiles (red dots) for the normal (upper) and
	abnormal modes emitted from	PSR J1326$-$6700.
	The blue solid line represents the fitting based on the combination of three
	Gaussian components (black dashed lines).
	Here the profiles have been scaled to the same height to show the change in
	the pulse location and shape.}
	\label{pic:profiles}
\end{figure}

The average pulse profiles for the normal and abnormal modes are shown in
Figure~\ref{pic:profiles}, where the pulse peaks are normalized to unity.
A combination of three Gaussian components is adopted to fit the observed
profiles. 
The best-fitted parameters determined using the Levenberg-Marquardt algorithm
\citep{Press+etal+1992} are presented in Table~\ref{tab:profiles} with R-square
of 0.996 and 0.998 for the normal and abnormal modes, respectively.
It demonstrates the existence of three emission components in the abnormal mode.
The relative amplitudes of the central and trailing components decrease with
respect to the leading component, and the three components shift earlier in
pulse phase.

\begin{table*}[h]
	\centering
	\caption{Parameters of the fitted Gaussian profile components for both
	emission modes from PSR J1326$-$6700, $I_i$ is the amplitude of the $i$th
	Gaussian component, $\phi_i$ is the peak longitude and $W_i$ is the full
	width at half peak.}
	\label{tab:profiles}
	\begin{tabular}{cccccccccc}
		\hline
		\hline
		Mode & $I_1$ & $\phi_1$ (deg) & $W_1$ (deg) & $I_2$ & $\phi_2$ (deg) & $W_2$ (deg) & $I_3$ & $\phi_3$ (deg) & $W_3$ (deg) \\
		\hline
		Normal & 0.21$\pm$0.01 & 157.30$\pm$0.07 & 2.6$\pm$0.1 & 0.802$\pm$0.004
		& 166.5$\pm$0.1 & 16.0$\pm$0.1 & 0.81$\pm$0.01 & 180.59$\pm$0.04 & 6.87$\pm$0.06 \\
		Abnormal & 0.53$\pm$0.01 & 155.62$\pm$0.03 & 3.50$\pm$0.05 &
		0.47$\pm$0.01 & 159.8$\pm$0.1 & 10.2$\pm$0.1 & 0.177$\pm$0.005 &
		168.8$\pm$0.5 & 27.4$\pm$0.4 \\
		\hline
	\end{tabular}
\end{table*}

\subsection{Mode fractions and timescale}
\begin{figure}[h]
	\centering
	\includegraphics[width=8.0cm,height=8.0cm,angle=0]{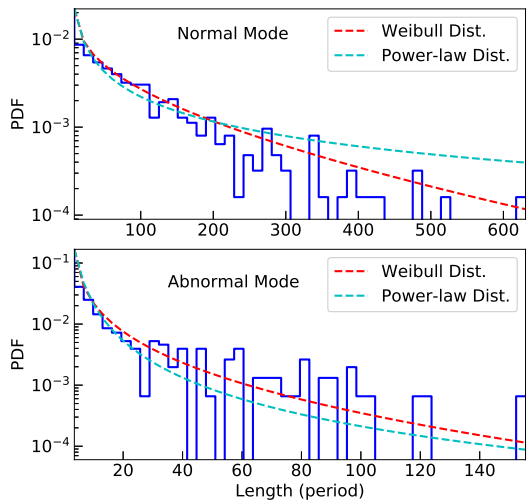}
	\caption{Duration distributions for the normal (top panel) and abnormal
	(bottom panel) modes taken from all observations together.
	The best-fit Weibull and power-law distributions for each mode are
	overdrawn.}
	\label{pic:duration_dist}
\end{figure}

In our observations, PSR J1326$-$6700 spends ~85\% of the time in the normal
mode, and ~15\% in the abnormal mode.
The probability density functions (PDFs) for the lengths of the normal and
abnormal modes taken from all observations together are 
presented in the upper and lower panels of Figure~\ref{pic:duration_dist}, respectively.
They show that the occurrence of both the modes decreases at longer timescales.
The PDFs of the two modes are well fitted by a Weibull distribution:
\begin{equation}
	\centering
	\rm
	P(\Delta t) = \frac{k}{\lambda} (\frac{\Delta t-\theta}{\lambda})^{k-1}
	e^{-(\frac{\Delta t - \theta}{\lambda})^k},
\end{equation}
where $k$ is the shape parameter, $\lambda$ is the scale parameter
and $\theta$ is the location parameter.
Considering the asymmetric distributions, the fittings are
performed on unbinned data by maximum likelihood estimation.
The best fitting coefficients are estimated to be
$k=0.79\pm0.03$, $\lambda=67\pm4$, $\theta=6.6\pm0.5$ and R-square=0.991 for the
normal mode, and $k=0.81\pm0.03$, $\lambda=11.8\pm0.7$, $\theta=2.81\pm0.04$ and
R-square=0.998 for the abnormal mode.
To validate the goodness-of-fit of the Weibull distribution
fits, a one-sample nonparametric Anderson-Darling test is carried out.
The durations of the normal and abnormal emission modes are both drawn from 
Weibull distributions at 75\% confidence level.
For both normal and abnormal modes, the Weibull distribution is
applied to all observations.
The same parameters are obtained at lower significance level, implying that
the underlying physical processes which produce the moding are fixed and unchanging.

The Weibull distribution is commonly used to analyze life data.
A $k<1$ implies that the probability of a mode change
occurring decreases with time, the shorter the pulsar is in a mode.
If $k=1$, an exponential distribution is produced, which indicates that the 
probability of a mode change occurring is time-invariant.
If $k>1$, the occurring of mode changing increases with time.
The best fitted power-law distributions for both modes are shown in
Figure~\ref{pic:duration_dist} as well, which do not fit the observed duration
distributions well.
For PSR B0919+06, the abnormal emission events occur randomly
about every 1000-3000 periods \citep{Han+etal+2016}, which is much longer than PSR J1326$-$6700.

\subsection{Mode-segregated polarimetric profiles}
\begin{figure*}[h]
	\centering
	\includegraphics[width=8.0cm,height=6.0cm,angle=0]{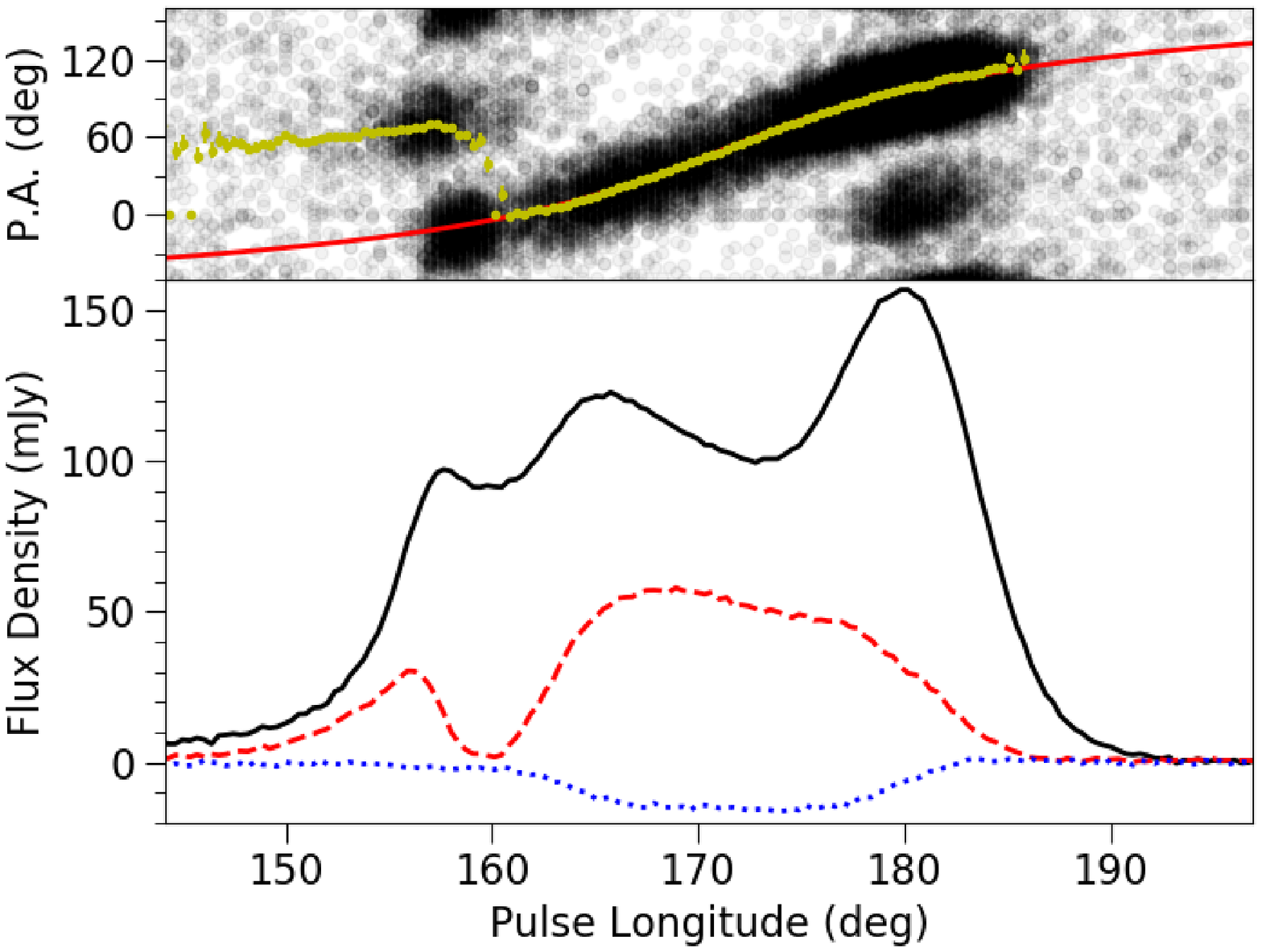}
	\includegraphics[width=8.0cm,height=6.0cm,angle=0]{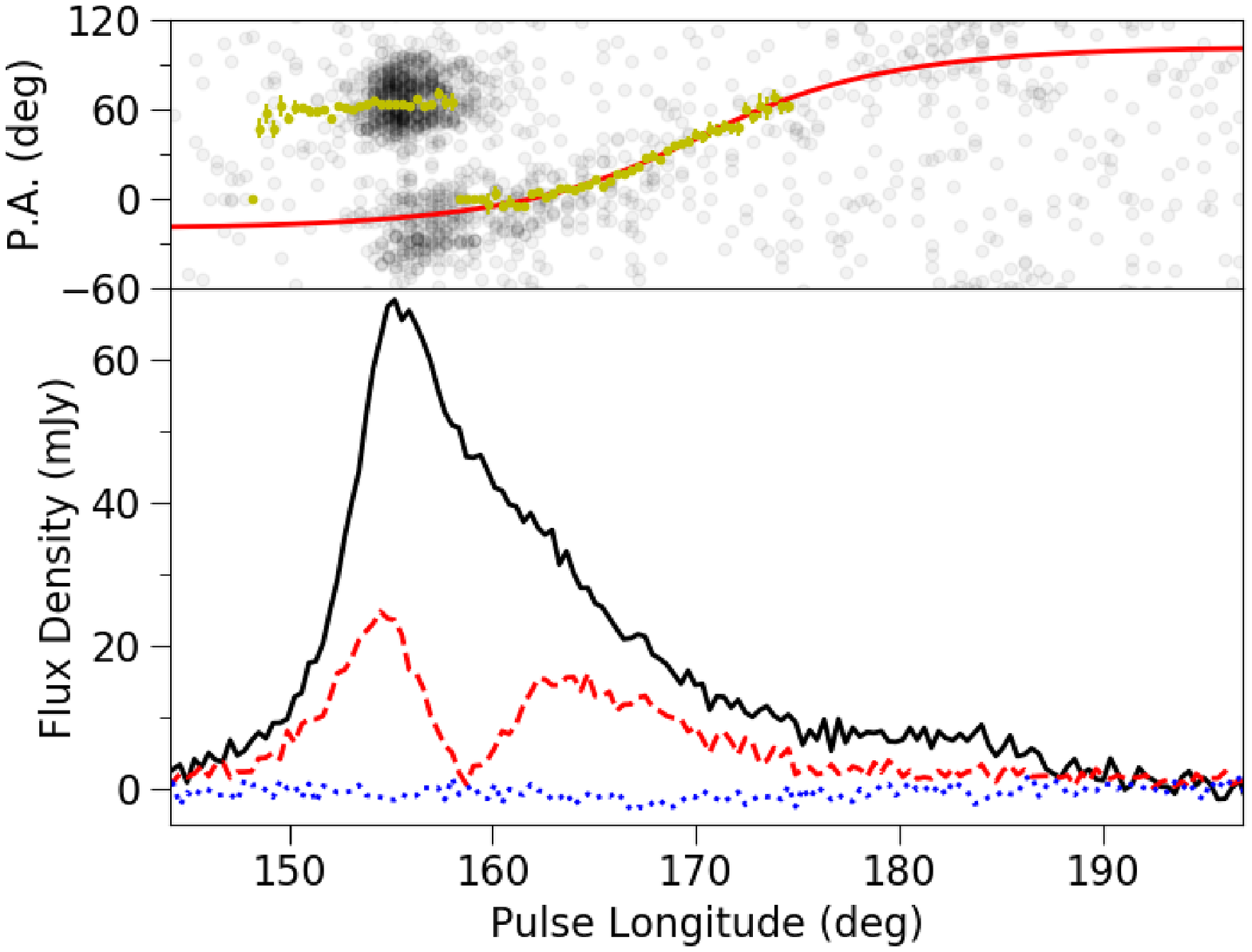}
	\caption{The averaged polarization profiles of normal (left) and abnormal
	(right) modes for PSR J1326$-$6700 at 1369 MHz.
	The lower panel gives the total intensity (Stokes I, black solid curve), the
	total linear (Stokes $L=\sqrt{Q^2+U^2}$, red dashed curve), and the circular
	polarization (Stokes V, blue dash-dotted curve).
	The upper panel presents the longitude-dependent histogram of measured 
	position angle $\psi=\frac{1}{2}tan^{-1}(U/Q)$ along with the position angle
	of the mean profile, a fitted curve computed based on the rotating vector
	model is shown in the red curve.}
	\label{pic:poln_profiles}
\end{figure*}

In order to understand the emission properties further, sets of normal and
abnormal integrated polarimetric profiles are constructed from 2012$-$01$-$15,
2014$-$05$-$25 and 2014$-$05$-$30 observations, and shown in
Figure~\ref{pic:poln_profiles}.
The bottom panels present the total intensity (black solid curves), total linear
polarization (red dashed lines) and circular polarization (blue dotted lines).
The upper panels give the linear polarized position angle (hereafter PA)
histograms for all single pulses, as well as the average PA traverses for normal 
and abnormal modes.
To limit the effects of measurement uncertainties, the PAs for each single pulse
are estimated at longitudes where the linear polarization power is more than 6
times the noise rms in the off pulse region.

The asymmetric total power profile of the normal mode has three main emission
components, with the trailing one the strongest.
During the abnormal emission state, however, the mean pulse profile is very
different in amplitude and pulse phase.
The emission from the normal profile ceases and the abnormal emission emerges at
a new leading component.
The abnormal profile possesses a single emission component with sharp leading
and gradual trailing edges, and the profile position shifts markedly earlier.
The mean flux density of normal mode is around 4 times brighter than that of
abnormal mode.
Table~\ref{tab:modes} lists the mean flux density, $W_{50}$ (width of pulse at
50\% of peak) and $W_{10}$ (width of pulse at 10\% of peak) of both normal and 
abnormal modes.
Note that the strong linear polarization under the trailing feature simply
disappears during the abnormal mode.
Significant right-circular polarization is observed in the center of the normal
profile.
The abnormal mode shows a marginal circular polarization over the whole profile.

\begin{figure}[h]
	\centering
	\includegraphics[width=8.0cm,height=6.0cm,angle=0]{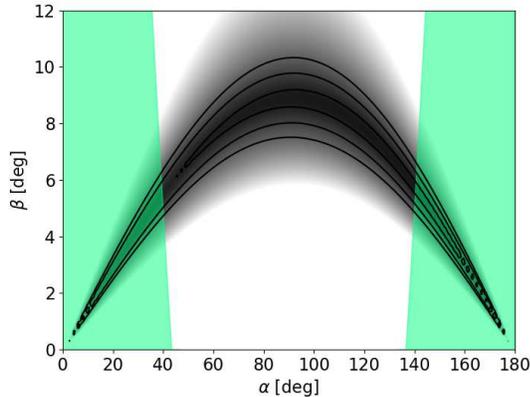}
	\caption{The results of fitting an RVM curve for each ($\alpha$, $\beta$)
	combination for normal mode.
	The reduced chi-square ($\chi^2$) of the fit is shown as the gray scale,
	with the darkest value corresponding to the best fit.
	Evidently, $\alpha$ and $\beta$ are highly correlated resulting in an
	ambiguity in the determination of $\alpha$.
	The black contour lines represent 1$-\sigma$, 2$-\sigma$ and 3$-\sigma$
	confidence boundaries.
	The green regions show constrained viewing geometries allowed by the
	observed pulse width.}
	\label{pic:alpha_beta}
\end{figure}

The PA exhibits a fast swing across the profile, and a `S'-shaped sweep is
presented, which may indicate that we are seeing the whole conal beam.
Then the line of sight geometry can be assessed by fitting the rotating vector 
model \citep[RVM,][]{Radhakrishnan+Cooke+1969} to the PA traverse.
It is noted that for both emission modes a jump of around of $60^\circ$ is shown
in the PA under the leading component which is also associated with a substantial
dip in the linear polarization.
For the normal mode, a bimodal distribution of PAs separated by around of
$80^\circ$ predominates in the trailing emission region as well.
The two preferred PAs at the leading and trailing components manifest the
presence of two orthogonal polarization states.
The classical RVM curve could be severely corrupted by the variations in
strength between the two modes.
In order to refrain the modal effects, only the pulse longitudes where the
average PA traverse is in good agreement with the single pulse PA in the central 
component are considered.
The best fits to the RVM are shown as red curves in the upper panels of
Figure~\ref{pic:poln_profiles}.
The viewing geometries for normal and abnormal modes are specified with the values of
magnetic inclination angle ($\alpha$), impact angle ($\beta$), position angle
offset ($\psi_0$) and fiducial plane angle ($\phi_0$), which are listed in 
Table~\ref{tab:modes}.
As shown that the quantities of $\phi_0$ and $\psi_0$ are significantly constrained.
However, the $\alpha$ and $\beta$ values obtained from these fits are extremely covariant 
and unreliable due to the limited duty cycle of the profile \citep{Mitra+Li+2004}.
For instance, Figure~\ref{pic:alpha_beta} shows the reduced $\chi^2$ values of the
fit as a function of $\alpha$ and $\beta$ for the normal mode.
Here, the best fit occurs where $\chi^2$ reaches a minimum.
Evidently, there are a number of combinations of $\alpha$ and $\beta$ which
provide equally acceptable fits.
Therefore, the actual geometry of the system can not be necessarily represented
by the derived angles.
From the RVM fit alone, we can conclude that $0^\circ<\beta<10^\circ$,
corresponding to a positive gradient of the PA swing.
Nevertheless, the fact that $\alpha$ is practically unconstrained.
More constraints on these parameters are described in the discussion section.

\begin{table*}[h]
	\centering
	\caption{List of the parameters during the normal and abnormal modes of PSR
	J1326$-$6700.
	The angles $\alpha$, $\beta$, $\phi_0$ and $\psi_0$ are fits obtained from
	the RVM.}
	\label{tab:modes}
	\begin{tabular}{ccccccccc}
		\hline
		\hline
		Mode & \% of pulses & Flux density (mJy) & $W_{50}$ (deg) & $W_{10}$
		(deg) & $\alpha$ (deg) & $\beta$ (deg) & $\phi_0$ (deg) & $\psi_0$ (deg) \\
		\hline
		Normal & 85\% & 11.40$\pm$0.01 & 27.9$\pm$0.2 & 37.0$\pm$0.5 & 179.5$\pm$4926.0
		& 0.08$\pm$782.45 & 171.3$\pm$0.2 & 50.5$\pm$0.9 \\
		Abnormal & 15\% & 2.44$\pm$0.01 & 10.4$\pm$0.3 & 37.3$\pm$0.7 & 0.07$\pm$35866.97
		& 0.01$\pm$5523.61 & 170.1$\pm$0.6 & 41$\pm$3 \\
		\hline
	\end{tabular}
	\begin{flushleft}
		\textbf{Notes.} The extremely high uncertainties in $\alpha$ and $\beta$
		reflect their covariant relationship.
	\end{flushleft}
\end{table*}

\subsection{Fluctuation spectra}
\begin{figure}[h]
	\centering
	\includegraphics[width=8.0cm,height=12.0cm,angle=0]{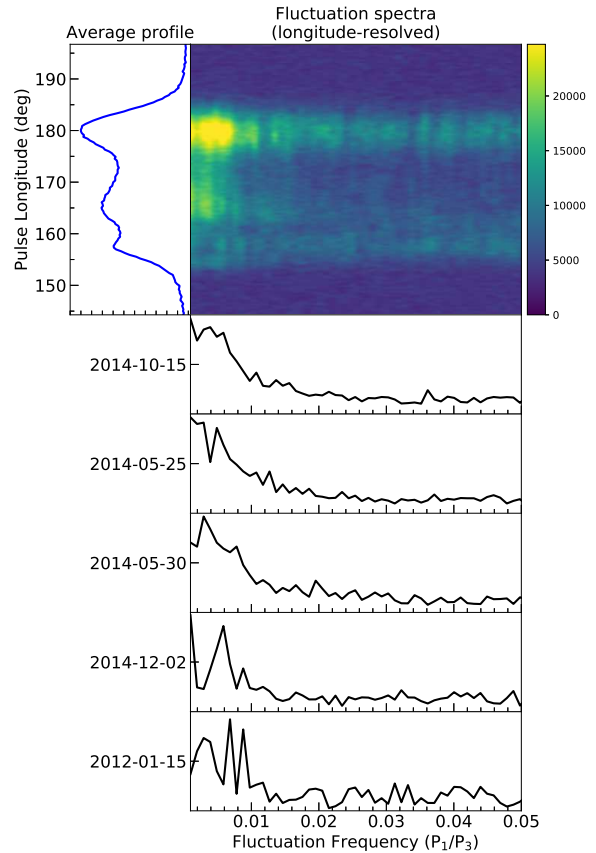}
	\caption{Fluctuation spectra of PSR J1326$-$6700, showing evidence for mode
	switching periodicity.
	The longitude-resolved fluctuation spectrum for 2014-10-15 is shown in the
	top right panel.
	The averaged pulse profile is shown in the left panel, and the lower panel
	displays the overall fluctuation spectrum.
	The following lower panels display the overall fluctuation spectra for
	2014-05-25, 2014-05-30, 2014-12-02 and 2012-01-15, respectively.
	Fourier transforms of length 1024 were used.
	Note that all of the overall fluctuation spectra are plotted only to 0.05
	cycles period$^{-1}$.
	The feature at 0.0039 cycles period$^{-1}$ corresponds to a periodicity of
	256 rotation periods.}
	\label{pic:lrfs}
\end{figure}

As shown in Figure~\ref{pic:sgl_demo}, the mode switching events seem to appear
such frequently with a rough regularity of several hundred
pulses in PSR J1326$-$6700.
The longitude-resolved fluctuation spectra \citep[LRFS;][]{Backer+1973} are calculated
to investigate whether the mode switching occurs with a regularity.
This involves performing discrete fast Fourier transforms along each longitude bin
within the pulse window.
Any periodicity will be indicated as a peak in the Fourier spectrum.
The spectrum of the 2014-10-15 observation of PSR J1326$-$6700, shown in the top 
panel of Figure~\ref{pic:lrfs}, remarkably displays a well-defined periodic feature 
with a peak frequency at 0.004 cycles period$^{-1}$ associated most strongly with the 
central and trailing components of the profile.
The latter property ties this principal fluctuation to the mode switching
events, since a longitudinal shift of emission would be most evident at these
longitudes.
A more accurate value for the total fluctuation frequency can be determined with
the peak value with an uncertainty estimated as FWHM/$\rm{2\sqrt{2ln(2)}}$, where 
FWHM is the full width at half-maximum of the peak feature and $\rm{2\sqrt{2ln(2)}}$
is the scaling for Gaussian approximation \citep{Basu+etal+2019}, giving 
$P_f=256\pm30$ rotation periods.
The total fluctuation spectra of all observations are shown in the following
lower panels of Figure~\ref{pic:lrfs}.
The somewhat long 2014-05-25 and 2014-05-30 spectra present period of $205\pm18$ and
$341\pm48$, respectively.
The short 2014-12-02 observation shows a strong feature at 0.0059 cycles
period$^{-1}$ with a period of $171\pm20$ stellar rotations.
Three strong peaks at 113, 147, and 333 periods are shown in the 2012-01-15 spectrum.
The somewhat different spectra presented maybe caused by the wide spreading of
fluctuation power in these observations.
Similar periodicities have also been identified in PSR B0919+06
with an approximate period of 150 rotation periods and PSR B1859+07 with a rough
period of 700 rotations \citep{Wahl+etal+2016}.
The quasi-periodicity found in the five separate fluctuation spectra 
is time variant, it is consequence of taking five separate observational samples
of unchanging Weibull distributions for the normal and abnormal modes.

In order to determine whether the periodic modulations are
originated from occasional phase drifting as suggested by \citet{Wang+etal+2007}, 
the two-dimensional fluctuation spectra \citep[2DFS;][]{Edwards+Stappers+2002} 
are calculated.
No evidence of drifting subpulses is shown in our observations.

\section{Discussion}
\label{sec:diss}

\subsection{The viewing geometry}
The location of the radio emission region in pulsar magnetosphere remains a
major uncertainty in our understanding of pulsar emission physics.
The emission altitude calculated using a geometrical approach involves a dipole
file configuration where the viewing geometry has to be specified
\citep{Phillips+1992}.
However, $\alpha$ and $\beta$ values cannot be unambiguously determined from the
RVM fitting as shown in Figure~\ref{pic:alpha_beta}.
An independent procedure with significant advantages for estimating emission
heights has been developed by \citep{Blaskiewicz+etal+1991}.
The steepest gradient point of the PA traverse is expected to have a time lag
with respect to the centroid of the total intensity profile due to relativistic
effects such as aberration and retardation.
Then the absolute emission height at the frequency of observation can be 
converted from this time delay ($\Delta \phi$), and is given by
\begin{equation}
	\centering
	\rm
	h_{em} = \frac{P c \Delta \phi}{8\pi},
\end{equation}
where $P$ is the rotation period of the star and $c$ is the speed of light.
The profile centroid $\phi_1$ is identified with the pulse longitude midway
between the outermost edges of the pulse intensity profile.
The uncertainty in $\phi_1$ is estimated by
$\sigma(\phi_1)=2\sigma(I)/|dI/d\phi|$, where $\sigma(I)$ represents the noise
level, and $dI/d\phi$ is the gradient of the profile in the vicinity of the
edge.
The pulse longitude at which the steepest slope of the PA curve occurs can be
measured at the fiducial plane.
The derived emission heights for normal and abnormal modes are given in
Table~\ref{tab:geometry}.

\begin{table}[h]
	\centering
	\caption{Profile centroid, beam width, $\rm{W_{open}}$ and emission altitude
	estimates for normal and abnormal modes.}
	\label{tab:geometry}
	\begin{tabular}{ccccc}
		\hline
		\hline
		Mode & $\phi_1$ (deg) & $r_{em}$ (km) & $\rho$ (deg) & $W_{open}$ (deg) \\
		\hline
		Normal   & 167.9$\pm$0.5 & 380$\pm$61 & 10.6$\pm$0.8 & 46.0$\pm$0.5 \\
		Abnormal & 161.1$\pm$1.4 & 1012$\pm$172 & 17.2$\pm$1.5 & 46.0$\pm$0.7\\
		\hline
	\end{tabular}
\end{table}

Subsequently, the half opening angle of the radio emission beam can be
determined from the emission height under the assumption that the beam is
bounded by tangents to the last open field lines of a dipolar magnetic field.
The half opening angle is obtained using the formula
\begin{equation}
	\centering
	\rm
	\rho = \theta_{PC} + arctan(\frac{1}{2} tan \theta_{PC}),
\end{equation}
where $\theta_{PC}$ denotes the angular radius of the open field line region,
and is given by 
\begin{equation}
	\centering
	\rm
	\theta_{PC} = arcsin(\sqrt{\frac{2 \pi h_{em}}{P c}})
\end{equation}
\citep{Lyne+Graham-Smith+2012}.

As a result, the values of $\alpha$ and $\beta$ can be constrained from 
\begin{equation}
	\centering
	\rm
	cos\rho = cos\alpha cos(\alpha+\beta) + sin\alpha sin(\alpha+\beta)
	cos(\frac{W_{open}}{2})
\end{equation}
\citep{Gil+etal+1984}, where $\rm{W_{open}}$ is the range of rotational
longitude for which the line of sight samples the open field lie.ne region.
$\rm{W_{open}}$ is taken to be twice the difference in phase between the
fiducial plane position and the pulse edge furthest from it
\citep{Rookyard+etal+2015}.
The green areas shown in Figure~\ref{pic:alpha_beta} show the favoured viewing
geometry that can produce a pulse of the measured width.
The magnetic axis is relatively aligned with the pulsar's rotation axis with
$\alpha<45^\circ$.

\subsection{Probing the mode switching mechanism}
As shown in Figure~\ref{pic:sgl_demo}, the time-dependent mode changing presents
gradual earlier shift pattern of the emission longitude, and occasional nulls
interrupt the abnormal emission state.
The $W_{10}$ remains constant for both emission modes. However, the $W_{50}$
during the abnormal mode narrows (see Table~\ref{tab:modes}).
Furthermore, in PSR J1326$-$6700 the profile changes occur quasi-periodically,
implying the existence of a different rotation frequency from that of the star.
It is suggested that the magnetosphere does not corotate with the star, and the
structure of the magnetosphere changes in a quasi-periodic pattern.
For instance, recently, the swooshes were identified with a quasi-periodicity in
pulsar B1859+07 and possibly B0919+06 \citep{Wahl+etal+2016}.

The origin of mode changing has been remaining a mystery since the first
discovery of mode switching in PSR B1237+25 \citep{Backer+1970}.
There is increasing evidence that the mode changing is possibly 
caused by changes of magnetospheric particle current flow \citep{Lyne+etal+1971}.
Nevertheless, the underlying physical mechanism for the change of magnetosphere 
state is not clear yet.
\citet{Lyne+etal+2010} presented the correlated changes in the pulse profile and
in the timing noise for six pulsars.
The switching between the radio-quiet `off' and radio-loud `on' states in the 
intermittent pulsar B1931+24 was found to be correlated with the slowing down 
rate of the pulsar \citet{Kramer+etal+2006}.
The abrupt changes of emission mode is suggested that the pulsar jumps between
two different magnetospheric states, and a causal connection between the radio
emission and the torque on the star is established physically.
A purely magnetospheric model for observed abrupt changes in pulsar radio
profile is applied to the swooshing in PSR B0919+06 \citep{Yuen+Melrose+2017}.
Therefore, the phase shift during the abnormal emission mode in PSR J1326$-$6700 
can be described by the shifts in the intersection points of the trajectory with 
the emission spot as well.
A long term monitoring is suggested to investigate whether the spin-down rate 
correlates with the shifting between the two profile states.

\begin{figure}[h]
	\centering
	\includegraphics[width=8.0cm,height=7.8cm,angle=0]{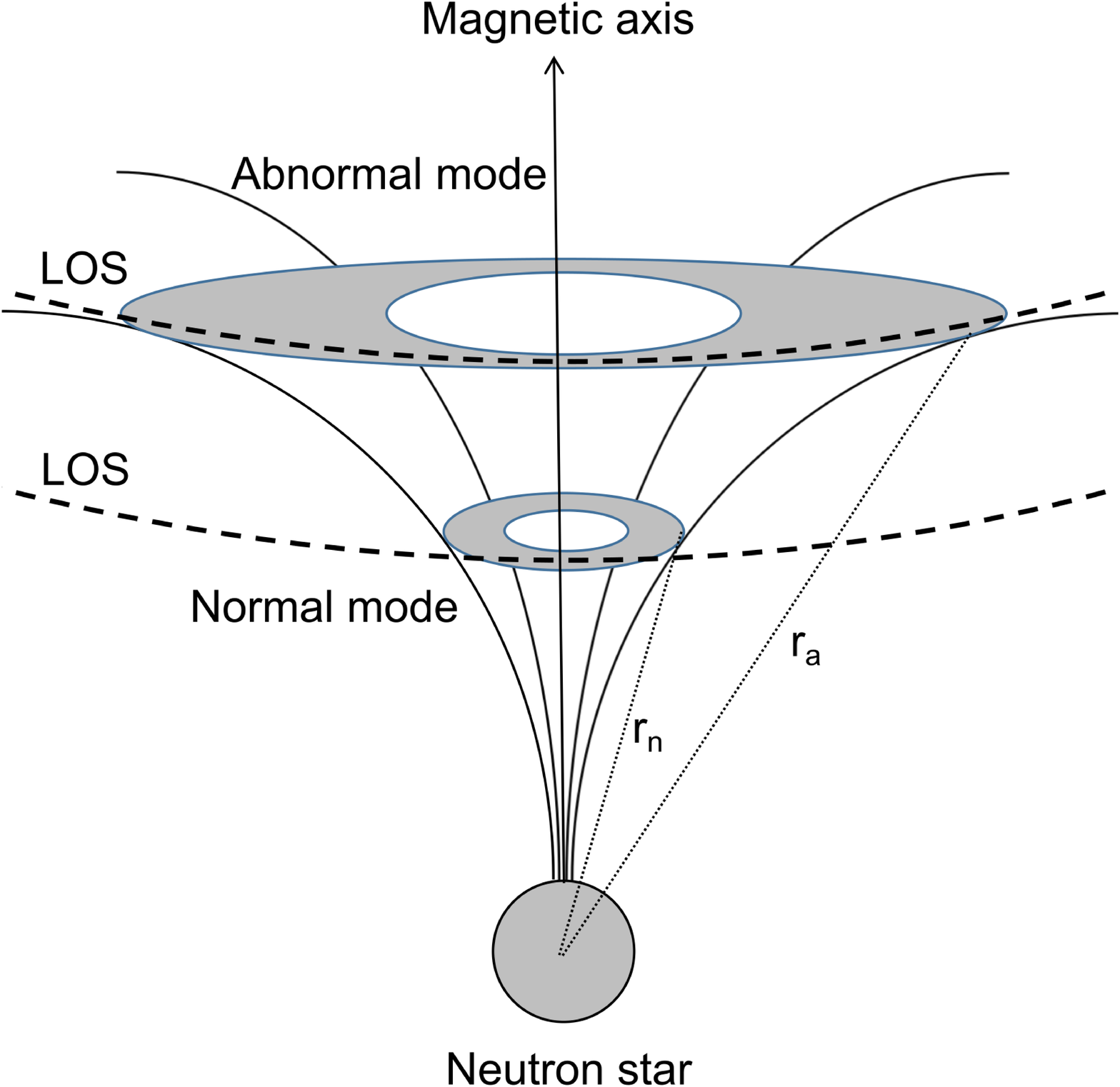}
	\caption{Schematic diagram explaining the observed behavior of two emission
	modes of PSR J1326$-$6700.
	The normal and abnormal emission regions, symmetrical around the magnetic
	axis, are assumed to be generated at different altitudes, $r_n$ and $r_a$,
	respectively.
	The beam size expands during the abnormal emission state, as shown by the
	trajectories of line of sight (LOS).}
	\label{pic:diagram}
\end{figure}

The delay emission heights for both modes derived using the relativistic beaming
model based on effects of aberration and retardation are given in
Table~\ref{tab:geometry}.
It is found that the abnormal emission arises at higher altitude from the
surface of the neutron star while the normal mode arise closer to the stellar
surface, corresponding to a jump from 0.015 to 0.04 light cylinder radius.
Furthermore, the beam opening angle of the abnormal mode is wider than that of
the normal mode.
A possible model is constructed, wherein the emission of both modes comes from 
the same magnetic flux surface,  but from different heights at a fixed frequency
\citep{Leeuwen+etal+2003}.
As shown in Figure~\ref{pic:diagram}, the altitude at a fixed frequency increases 
when the emission changes from normal mode to abnormal mode.
According to the traditional vacuum gap model \citep{Ruderman+Sutherland+1975}, 
the transition from normal mode to abnormal mode occurs by an increase in height 
of the voltage gap.
As a result the altitude of emission increases.

Alternative possible origins of the anomalous variations in the
on-pulse phase have been proposed.
\citet{Rankin+etal+2006} suggested that the swooshes could be caused by `partial 
conal' emission, where the emission region was partly obscured.
The binary companions in the light cylinder orbits was adopted to interpret the
inter-swoosh quasi-period \citep{Wahl+etal+2016}.

\section{Conclusions}
\label{sec:con}

PSR J1326$-$6700 presents two emission modes based on the distribution of
$\Delta \chi^2$ in our work.
Usually, the emission comes from three regions of the profile, but occasionally
the emission weakens at the central and trailing components along with shifts of
the leading emission towards earlier to illuminate the leading edge of the
profile for less than a minute.
The overall intensity of the normal mode is almost five times that of the
abnormal mode.
The $\Delta \chi^2$ distribution of the normal mode is wider than that of the
abnormal mode, which suggests that the normal emission mode is less stable than
the abnormal mode.
The pulsar spent 85\% of the total observation time in the normal mode, and 15\%
in the abnormal mode.
The durations of both emission modes can be well described by Weibull
distributions with shape parameters less than 1, which indicates that the
occurrence of mode changing is clustered.
Furthermore, a quasi-periodicity was found in the mode switching in pulsar
J1326$-$6700.

The high S/N profiles have allowed us to estimate the magnetospheric height of
the 1369 MHz emission, based on the delay-radius relation and the lag between
the phase at the steepest PA gradient and that at the profile's mid-point.
Using the observed phase lags, the emission height of the abnormal mode is
estimated to be around three times higher than that of the normal mode.
It is evident that the simultaneous multi-wavelength polarization observations
are necessary to offer further insights into the frequency evolution of mode 
changing in PSR J1326$-$6700 and to gain a full description of the physical
processes driving the changes.
For instance, why such a emission altitude change should occur and why the
height should increase rather than decrease during the abnormal mode.

\section*{Acknowledgements}
\addcontentsline{toc}{section}{Acknowledgements}
We are grateful the referee for helpful comments.
Much of this work was made possible by grant support from the Chinese National 
Science Foundation Grant (U1838109, U1731238, U1831102, U1631106, 11873080),
the West Light Foundation of Chinese Academy of Sciences (WLFC 2016-QNXZ-B-24),
and the National Basic Research Program of China (973 Program 2015CB857100).
J.P.Y. is supported by a prospective project of the Astronomical Research Center of
the Chinese Academy of Sciences.
N.W. is supported by the National Program on Key Research and Development Project
(grant No. 2016YFA0400804).
H.G.W. is supported by the 2018 project of Xinjiang uygur autonomous region of China 
for flexibly fetching in upscale talents.
J.L.C. is supported by the Scientific and Technological Innovation
Programs of Higher Education Institutions in Shanxi (grant No. 2019L0863).
The project was supported by Open Fund of Guizhou Provincial Key Laboratory of 
Radio Astronomy and Data Processing.
We thank members of the Pulsar Group at XAO for helpful discussions.
The Parkes radio telescope is part of the Australia Telescope National Facility, 
which is funded by the Australian Government for operation as a National 
Facility managed by CSIRO.

\bibliographystyle{aasjournal}
\bibliography{bibtex}

\end{document}